\documentclass[prd,aps,floats,12pt]{revtex4}
\usepackage{amsmath}
\usepackage{amssymb}
\usepackage{epsfig}
\def\d3h{\lower-1pt\hbox{$\stackrel{\ldots}{H}$}}

\def\beq {\begin{equation}}
\def\eeq {\end{equation}}
\def \ber {\begin{eqnarray}}
\def \eer {\end{eqnarray}}

\def \lleq {\lower0.9ex\hbox{ $\buildrel < \over \sim$} ~}
\def \ggeq {\lower0.9ex\hbox{ $\buildrel > \over \sim$} ~}
\def\ie {{\it ie~}}
\def\etal {{\it etal.~}}

\def \z {Zeldovich~}
\def \s {Starobinsky~}
\def \za {Zeldovich approximation~}

\begin {document}

\title{Ya.~B.~Zeldovich (1914-1987) \\Chemist, Nuclear Physicist, Cosmologist}

\author{Varun Sahni}\email{varun@iucaa.ernet.in}
\affiliation{Inter-University Centre for Astronomy and Astrophysics,
Post Bag 4, Ganeshkhind, Pune 411~007, India} %

%\centerline{\bf Ya.~B.~Zeldovich (1914 --  1987)}
%\centerline{A brief biographical sketch and some personal reminiscences.}

\begin{abstract}
\bigskip

Ya.B. Zeldovich was a pre-eminent Soviet physicist whose seminal contributions
spanned many fields ranging from physical chemistry to nuclear and particle physics,
and finally astrophysics and cosmology. March 8, 2014 marks Zeldovich's birth centenary, and this 
article
attempts to convey the zest
with which Zeldovich did science, and the important role he played in fostering and mentoring
a whole generation of talented Scientists.
%(Written for the Indian popular science journal Resonance.)

\end{abstract}

\maketitle

\section{Introduction}

Yakov Borisovich Zeldovich was very talented. His active scientific
career included major contributions in fields as diverse as chemical physics
(adsorption \& catalysis),
the theory of shock waves,
thermal explosions, the theory of flame propogation,
 the theory of combustion \& detonation, 
nuclear \& particle physics, and, during the latter part of his life:
gravitation, astrophysics and cosmology \cite{zel}.

\z made key contributions in all these area's, nurturing a creative and
thriving scientific community in the process.
His total scientific output exceeds 500 research article and 20 books.
Indeed, after meeting him, the famous English physicist Stephen Hawking wrote 
``Now I know that you are a real person and not a group of scientists like
the Bourbaki''.$^1$\footnotetext[1]{Bourbaki was the pseudonym collectively adopted by
a group of twentieth century mathematicians who wrote several influential books under
this pseudonym on advanced mathematical concepts.}
 Others have compared his enormously varied scientific output
to that of Lord Raleigh who, a hundred years before \z,
 worked on fields as varied as optics and engineering.

Remarkably \z never received any formal university
education !
He graduated from high school in St. Petersburg
 at the age of 15 after which he joined the {\em Institute for Mechanical Processing
of Useful Minerals}
(`Mekhanabor') to train as a laboratory assistant.
The depth of \z's questioning
and his deep interest in science soon reached senior members of the
scientific community and, in 1931, the influential soviet scientist
A.F. Ioffe wrote a letter to Mekhanabor requesting that \z be ``released
to science''. \z defended his PhD in 1936 and, years later, reminiscenced of the
``happy times when permission to defend [a PhD] was granted to people who had
no higher education''.

Despite his never having been formally taught (or perhaps because of it !)
\z developed a very original style of doing science,  and became, in the process,
an exceptional teacher. It is also interesting that in his early years 
\z had been an experimentalist
as well as a theoretician, and this closeness to complementary aspects
of science guided him throughout his later life.

\section{Nuclear and Particle Physics}

It is quite remarkable that \z ventured into a totally new field --
Astrophysics -- around 1964 when he was nearing 50.
%This was quite unusual, since 
By then \z had already developed a very 
considerable reputation in
fields ranging from physical chemistry (there is a `Zeldovich
number' in combustion theory) to nuclear physics. 
Indeed, having done pathbreaking work on
combustion and detonation, \z moved to nuclear physics in the 1930's 
writing seminal papers  demonstrating the
possibility of controlled fission chain reactions among uranium isotopes.
This was the time when fascism was on the rise in Germany, and,
in an effort for national survival, the Soviet Union
was developing its own atomic program of which \z (then in his mid 20's)
quickly became a key member. 
According to Andrei Sakharov, ``from the very beginning of Soviet work
on the atomic (and later thermonuclear) problem, \z was at the very epicenter of
events. His role there was completely exceptional'' \cite{sakharov}.
One might add that \z's earlier work on combustion paved the way for creating the
internal ballistics of solid-fuel rockets which formed the basis of
the Soviet missile program during the `great patriotic war' and after \cite{ginzburg}.
Sadly, much of Zeldovich's work during this period remains classified
to this day.

After the war \z went on to do pioneering work in several
 other aspects of nuclear and particle physics including
%A small subsample of his major contributions is
%summarized below 
$^2$\footnotetext[2]{Space does not permit me to elaborate on \z's other seminal
work in this area which included the possibility of muon-catalyzed fusion (1954),
his work on weak interactions including his brilliant hypothesis
on the existence of parity violating neutral currents (1959), his remark (1957)
 about
the existence of
a toroidal dipole moment (which has since spawned an active research area)
and his prediction about the existence of the $\eta$ meson (1958) which was subsequently
discovered.
The reader is referred to \cite{zel} for more details.}:

\begin{itemize}

\item In 1952 and 1953 \z proposed laws for the conservation of lepton and
baryon charges.

\item In 1955 \z and Gershtein suggested the conservation of the weak charged
vector hadron current. This idea was independently discovered some years later by
Feynman and Gell-Mann and played a key role in the development of the theory
of weak interactions.
%put forth a very important hypothesis on the existence of
%neutral currents which violated parity conservation. 
\z (1959) also suggested the existence of neutral currents which violated parity conservation.
He showed
that parity violation in weak interactions should lead to the rotation
of the plane of polarization of light propogating in a substance not containing
optically active molecules. This prediction was subsequently confirmed.

\item In 1959 \z suggested a method of containment (storage) of slow neutrons
via total internal reflection from graphite. This method is now regularly used
to measure the neutron electrical moment.

\item \z also studied the possible existence of long-lived nuclei with
a large isotopic spin. He suggested the possibility of observing an isotope
of $^{8}He$, soon after which this isotope was, in fact, discovered !

\end{itemize}

\section{Astrophysics and Cosmology}

Instead of resting on his laurels 
(had he so desired \z could have easily landed a `comfortable' job heading
a premier research laboratory
or institute), \z decided to change course midstream and, from about 1964,
devoted his phenomenal abilities to problems in astrophysics.
$^3$\footnotetext[3]{
By this time \z was a very well known
figure in Soviet science and had won several major laurels including
three gold stars, the Lenin prize and a fellowship of the prestigious
Soviet Academy of Science.}
A change of track can be precarious
if made later in life when ones scientific tastes and habits
have usually become set.
Indeed, each scientific discipline has its own nuances the mastering of which can 
take considerable time and effort, and
history is replete with examples of scientists -- exceedingly capable
in their chosen field -- making grave errors of judgement when moving to another.

It is therefore quite remarkable that \z not only plunged deeply
into astrophysics, he virtually 
revolutionized the field, becoming in the process
one of the founders of relativistic astrophysics and physical 
cosmology. %(The other Russian scientist who contributed greatly to the
%development of modern cosmology was George Gamow.)

Below is an imperfect attempt to summarize some of Zeldivich's seminal 
contributions to astrophysics and cosmology.

\begin{itemize}

\item In 1963 \z and Dmitriev %derived the following elegant relationship between
%the kinetic energy ($K$) and potential energy ($U$) of a system
showed that the total energy of a system of particles interacting via gravity
 (in a universe expanding according to Hubble's law ${\bf v} = H{\bf R}$)
evolved according to the simple formula
\beq
\frac{dE}{dt} = -(2K+U)H~,
%\frac{d}{dt}(K+U) = -(2K+U)H~.
\label{zi}
\eeq
where $E=K+U$ is the total energy of the system -- $K$ and $U$ being
its kinetic and potential
energy respectively.
%This relationship can be used to estimate the mean gravitational binding energy associated
%with `dark matter' 
(This relation was independently suggested by Layzer (1963) and Irvine (1961)
 and is frequently referred to as the Cosmic Energy Equation.)
%to as the Layzer-Irvine relation in the western literature on cosmology.)
${\dot E} = 0$ once matter decouples from cosmological expansion.
In this case (\ref{zi}) reduces to the virial equation $2K+U=0$ which can be used to
investigate the amount of dark matter associated with gravitationally bound systems
such as clusters of galaxies.

\item \z was intrigued by black holes and spent a considerable amount of effort
in understanding their properties.  In 1962 \z published  a paper in which he
 showed that a black hole could be formed not only during the course
of stellar explosions, as was then widely believed, but by any mechanism which compressed
matter to sufficiently high densities. This opened up the possibility of
the formation of microscopically small black holes in the early Universe.
Were they to survive until today, the smallest holes would have masses of about
$10^{16}$ grams, roughly that of a large mountain.$^4$\footnotetext{Smaller mass black holes
would have evaporated by emitting Hawking radiation and disappeared.}
There has also been some speculation that microscopic black holes
may be the {\em dark matter} that everyone is searching for !
(This was Zeldovich's first paper on General Relativity. 
It was also the last work that he 
discussed with his teacher, Lev Landau,
before the latter's tragic car accident in 1962.)
%who met with a serious car accident a few days later.)

\item  In 1964 \z suggested that a black hole may be detected by its influence
on the surrounding gas which would accrete onto the hole. 
(The same result was independently obtained by E. Salpiter 
in the USA.) In 1966 \z also suggested (with Guseinov)
that one could look
for a black hole in binary star systems through the holes influence on
the motion of its bright stellar companion. 
These papers helped create a paradigm shift in which black holes were 
elevated from their earlier status of `impossible to observe passive objects'
to {\em objects which created very significant astrophysical activity in their
vicinity}.
%impossible to observe to 
%an awareness into the possibility that a
%black hole, though formally invisible, may nevertheless be detected
%indirectly through its influence on its surroundings.
Thus Zeldovich's early work led to what is currently a thriving area in 
astrophysics, with numerous 
observational programmes being dedicated to the discovery of
black hole candidates in our own galaxy as well as in distant galaxies and QSO's.
%have been detected using the methods suggested by \z.

\item \z remained deeply interested in black holes and, in 1971, turned his attention
to the issue of particle production and vacuum polarization in the 
strong gravitational fields that one would expect to encounter near
black holes and during the early infancy of the universe.
\z (1971) and his student Starobinsky (1973)
 showed that under certain conditions
a rotating black hole could loose energy via the production of a
particle-antiparticle pair. %In 1971 \z and Starobinsky wrote a seminal
%paper demonstrating that particles and antiparticles
%could have been copiously produced in the early universe.
%from the vacuum could have been very copious in the early universe.
%exceeding in some cases, the density of `normal' matter and radiation.
These papers were precursors of a whole body of later work including Stephen Hawking's
famous paper on evaporating black holes published in 1975.
(Hawking has on several occasions acknowledged his appreciation of the earlier work of
the two Russian scientists.)

\item In tandem with their study of quantum effects near black holes,
\z and Starobinsky began a systematic investigation into quantum effects
which occur
during the early stages of the universe when it was expanding
very rapidly.
Quantum field theory informs us that, far from being empty,
 the vacuum is actually seething with activity in the form of
virtual
particle-antiparticle pairs in the constant process of creation and
distruction. From the uncertainty principle 
$\Delta E \Delta t \geq \hbar$ we know that such pairs 
(of mass $2m$) come into existence for a very short time 
$\Delta t \leq \hbar/2mc^2$. 
If an electric field is applied to the vacuum then, 
for a sufficiently large value ($E_{\rm cr}$), the work done ($\delta W$)
on the virtual pair can become equal to the total rest mass $2mc^2$: 
\beq
\delta W = \lambda_c\vert e E_{\rm cr}\vert \simeq 2mc^2~,
\eeq
where the Compton length $\lambda_c = \hbar/mc$  provides an estimate of
the separation between particle and antiparticle. The critical field
value at which the vacuum becomes unstable to particle-antiparticle
production, $E_{\rm cr} \simeq m^2c^3/\vert e\vert \hbar$,
is called the Schwinger limit, after Julian Schwinger who discovered
it in 1951. ($E_{\rm cr} \sim 10^{16} ~{\rm volt/cm}$ for electrons.)

\z and \s showed that a similar effect occurs if the universe is
expanding rapidly. In this case the role of the electric field
is played by the gravitational field
which, in Einsteins general relativity,
is an expression of space-time curvature. %In other words,
A rapidly expanding universe literally tears
a virtual particle-antiparticle pair apart, giving rise to spontaneous
particle creation from the vacuum. Near the big bang the universe is expanding
very rapidly, (its rate of expansion being given by the Hubble parameter
$H = {\dot a}/a$), %which formally becomes infinite at the big bang itself !
and copious particle production from the vacuum is expected when
the expansion rate becomes of order the particle mass ($H \sim m$). 

In an influential paper published in 1971, \z and \s showed that particle 
production from the vacuum
becomes extremely significant if the universe expands 
{\em anisotropically}~$^5$.
\footnotetext[5]{The present-day universe expands according to Hubble's law
${\bf v} = H{\bf r}$, where the Hubble parameter, $H$, is a scalar
quantity whose value depends only upon time but not upon spatial 
direction. During anisotropic expansion
the expansion rate is different along the three spatial
directions so that %$v_x = H_1 x, v_y = H_2 y, v_z = H_3 z$ or,
%more generally, 
$v_i = H_{ij}r^j$, with the Hubble parameter being promoted to a tensor.} 
In this case the copious energy density
of particles released from the vacuum 
far exceeds that of any pre-existing matter in the universe !
Furthermore, the newly created particles (and their anti-particles) backreact on
the universe through the semi-classical
 Einstein equations, $G_{ik} = 8\pi G \langle T_{ik}
\rangle$,
isotropising its expansion.
The mechanism proposed by \z and \s provided an interesting means of
making the properties of the universe similar to what we
observe today. 
It is well known that the most general solutions to the Einstein equations 
will be both inhomogeneous and anisotropic. 
%ameliorated a long standing problem in cosmology. In 
%1973 Collins and Hawking had shown that a universe which resembled our own, 
%and appeared both homogeneous as well as isotropic 
%on large scales, is mathematically extremely improbable
%(`a set of measure zero').
Even if one were to restrict oneself to homogeneous models
(whose spatial properties did not depend upon the precise location
of the observer), a universe which expanded at different rates along
different directions was, in a sense, much more likely than our own
isotropic universe. This dilemma was substantially ameliorated by the
work of \z and \s  since copious particle production (and vacuum
polarization) would ensure that a universe whose expansion was initially
anisotropic soon isotropized
and began to resemble our own ! %soon after the big bang ! 
%Thus the
%set of initial conditions from which our universe could emerge was 
%considerably expanded: the quantum mechanism ensuring that 
%soon after the Big Bang the universe
%might begin to resemble our own !%\footnote{A similar result -- namely the rapid
%isotropisation of the universe -- %was later shown to hold for
%also holds for a space-time filled with a matter
%having the unusual equation of state $P \simeq -\rho$.
%This result, known as the cosmic `no hair theorem', was
%discovered by several people including the author.}
% a cosmological 
%which expanded exponentially fast during an early `Inflationary'
%epoch.

%\z subsequently ascribed to the vacuum a property which he called 
%`vacuum viscosity'. 

\item In 1966 \z and Gershtein showed that,
if neutrino's were massive, then they could very easily be the 
dominant matter component in the universe.
The reason has to do with the fact that the theory of weak interactions 
predicts a relic abundance for neutrino's of roughly 100 particles per cubic 
centimeter (per species). (By comparison the cosmic microwave background contains
 $\sim 400$ photons per cubic centimeter, also of relic origin.)
%which pervade the universe.)
If neutrino's were massive then, for a large enough mass, their density could
 easily exceed 
that of visible matter in the universe !
\z and Gershtein placed a limit on the
mass of the muonic (and electronic) neutrino from cosmological considerations.
Their result, $m(\nu_\mu) < 400~ ev/cm^3$,
 was considerably lower than laboratory 
bounds at the time, and convincingly demonstrated that the universe could
be used as a particle physics laboratory.
% arose from the fact that a larger neutrino mass
%would give rise to a smaller age for the universe, in contrast to observations
%which showed that the universe was at least 10 billion years old.
Subsequently Schwartzmann, a student of \z, showed how the number of neutrino
species could be constrained from observations of the helium abundance in the 
universe. (The reason is simple: a larger number of neutrino's speeds up cosmic expansion
and, in so doing, alters the primordial nucleosynthesis of light elements
taking place during the first few minutes of the big bang.)
These early papers by \z and his students proved to be prescient in defining 
a vibrant new field, {\em Astroparticle physics},
 in which the early universe plays the role of a particle-physics laboratory.
In the 1970's, %Vera Rubin and others demonstrated
the existence of a large amount of
 dark matter in galaxies was discovered observationally. 
\z's earlier work 
demonstrated that relic non-baryonic particles left over from the Big Bang 
could easily play this role !
Non-baryonic dark matter is currently believed to constitute
roughly a third of the total matter density in the universe --
significantly more than the $\sim 4\%$ contributed by baryons and electrons.$~^6$\footnotetext[6]{The possibility of massive neutrino's playing the role
of dark matter was pointed out by Cowsik \& McClelland (1973)
and, independently, by Marx \& Szalay (1972). However it is currently
believed that neutrino's account for only a small fraction
of the total dark matter density in the universe.}

\item \z returned to the issue of relic abundances 
a few years later in a seminal work on field theory.
Cosmology, like other disciplines,
 has frequently been influenced by developments in
neighboring fields. An example is provided by field theories including those in which
the ground state is degenerate so that the system can settle into different
ground states in different regions of space. (In ferromagnetism,
at temperatures below the Curie point, the magnetization vector
can point in any given direction. The phase transition in this case is of second order.)
\z investigated cosmological consequences of phase transitions which
could have occured in the early universe when its temperature was exceedingly high.
He demonstrated that theories with degenerate ground states
predicted a relic abundance of `new objects', called topological defects,
 which can be zero-dimensional -- such as magnetic monopoles,
one-dimensional -- such as cosmic strings, or even two-dimensional domain walls.
Topological defects arise during
an early phase transition once the universe has cooled below a critical value.
Their properties are similar to the
defects seen in laboratory physics such as vortices in a superfluid and
flux tubes in a superconductor. \z, Kobzarev and Okun (1974), and independently Kibble (1976),
appreciated the enormous impact that stable topological defects could have within
a cosmological setting.
\z and his colleagues demonstrated that whereas
monopoles and walls were disastrous for cosmology, cosmic strings might be useful
since they would act as `seeds' onto which matter accreted resulting in the formation
of galaxies and other gravitationally bound systems %including clusters and 
%superclusters of galaxies 
$^7$\footnotetext[7]{(i) Zeldovich's paper (with Khlopov) demonstrating that the
abundance of 't Hooft-Polyakov monopoles was unacceptably large appeared
in 1978, a year before the famous paper by Preskill which highlighted the same
problem but within the framework of grand unified theories.
(ii) Cosmic strings created during a grand unified
phase transition will have a radius much smaller than that of a proton,
an enormous length (upto a million light years) and a very high density:
a kilometer long cosmic string can be as massive as the earth !
These enormously long `one dimensional'
objects should not be confused with the much smaller
superstrings. }.

\item In 1967 \z applied himself to the issue of the cosmological constant `$\Lambda$'.
Originally introduced by Einstein in 1917, the cosmological constant
has the unusual property that, unlike other forms of matter, 
its pressure is {\em negative} and equal, in absolute terms, to its density
($P = -\rho$). 
Hence, while the density in normal
 forms of matter declines in an expanding universe, the density in $\Lambda$
remains frozen to a constant value $\rho = \Lambda/8\pi G$. After its inception the
 cosmological constant had fallen
into disrepute since it did not seem to be required by observations. Even Einstein
distanced himself from it, calling the $\Lambda$-term `my biggest blunder'.
\z radically changed this perspective by persuasively arguing
that, within the context of quantum field theory,
the prospect of a non-zero value for the $\Lambda$-term should be taken
extremely seriously. 
The reason is that the quantum polarization of the vacuum results in a 
{\em vacuum energy} which, quite remarkably,
 has the precise form of a cosmological constant:
$\langle T_i^k\rangle = \Lambda\delta_i^k/8\pi G$. 
(The equation of state $P = -\rho$ is, in fact, Lorenz invariant and remains the
same in any coordinate system. So the properties of the vacuum do not change
from one coordinate system to another, which is indeed a desirable property.)
A positive cosmological constant can result in an accelerating
universe, and by strongly supporting the $\Lambda$-term, \z paved the way for future
advances including cosmic inflation in the 1980's and dark energy
in this century \cite{zel1}.  These new results suggest that the universe
accelerated both in its remote past (inflation) and at present (dark energy). 
Thus current observations
 appear to require a form of matter whose properties are
tantalizingly similar to that of the $\Lambda$-term and one is reminded of
Zeldovich's prescient statement, regarding $\Lambda$, made almost half a century ago  ``the genie has been let out of 
the bottle, and it is no longer easy to force it back in'' \cite{zel_lambda}.

\item The discovery of the cosmic microwave background (CMB) in 1964 resulted in \z
becoming an ardent believer in the hot big bang model of the universe
(originally proposed by another gifted Russian physicist George Gamow).
During 1968 -- 1971,
working with a dedicated team of students and researchers (including
Doroshkevich, Novikov and Sunyaev), \z showed 
how the CMB could be used to probe the properties of the early universe.
His work in this field focussed on several key  issues including that
of the cosmological
recombination of hydrogen from free protons and electrons
$^8$\footnotetext[8]{This took place about 100,000 years
after the big bang, when the temperature of the universe had dropped to 4000$^\circ$K. 
Prior to this the universe had been opaque to the passage of light.},
the nature of angular fluctuations in the CMB, %including the importance of
%the dipole and quadrupole as well as the damping of small scale fluctuations,
%due to the non-instanteity of the recombination process.
and how these could be linked to fluctuations in matter.
He also addressed
the problem of cosmological nucleosynthesis in the hot early universe.
%The presence of fluctuations in the CMB detected by the COBE satellite in 1992
%marked a turning point in our understanding of the universe 
%since it showed that the infant universe had not been featureless and smooth, 
%but had tiny fluctuations (approximately 1 part in 100,000) imprinted on it.
%The (fourier) spectrum of these fluctuation
%turned out to have the `scale-invariant form' suggested by \z more than two
%decades earlier. (These tiny fluctuations in matter become galaxies
%several billion years later, after
%being amplified by gravitational instability.)

\section{The Zeldovich approximation and the Cosmic Web}

\item \z himself felt that his main contribution to cosmology was in
the understanding of how gravitational instability develops 
in the universe
from small initial values.

A realistic model of galaxy formation clearly requires two essential
ingredients:

(i) A description of the initial fluctuations in the distribution of matter
 which might later form galaxies.

(ii) A theory describing the growth of these fluctuations under the influence
of gravity or other forces.

%Since (i) was well described by his approximation (\ref{za}), \z
%turned his efforts to an understanding of initial conditions.
%The resulting formalism which he published in 1972 was that initial
In 1972 \z published a paper which subsequently proved to be rather prescient.
In it he discussed how
`seed' fluctuations (which later gave rise to galaxies) could have a
{\em scale invariant} spectrum. Mathematically this meant that the
fourier amplitudes of the spatially fluctuating gravitational potential
had the form $|\phi(k)|^2 \propto k^{-3}$, so that
$\int |\phi(k)|^2d^3k = \int d\log{k}$. In other words, each logarithmic
interval contributed an equal amount of power to the gravitational potential
responsible for moving matter into galaxies
$^9$. \footnotetext[9]{Additionally, if the phases of $\phi(k)$ are randomly distributed,
%in the interval $\lbrace 0, 2\pi\rbrace$ 
then the spatial
distribution of the gravitational potential, $\phi({\bf x})$, has the properties
of a Gaussian random field.}
%(It should be
%mentioned that 
(The scale invariant spectrum was independently discovered by
E. Harrison.) A decade after it was initially proposed on 
phenomenological grounds, the scale invariant spectrum was shown to
be a generic prediction of {\em Inflationary models} of the early Universe
where its origin was quantum mechanical.
%generate a spectrum similar to that suggested
%by Harrison and \z. 
%In 1992 the COBE satellite
%discovered fluctuations in the cosmic micrwave
%background having an a spectrum similar to that suggested
%by Harrison and \z. 
The presence of fluctuations in the CMB detected by the COBE satellite in 1992
marked a turning point in our understanding of the universe
since it showed that the infant universe had not been featureless and smooth,
but had tiny fluctuations (approximately 1 part in 100,000) imprinted on it.
The (fourier) spectrum of these fluctuation
turned out to have the `scale-invariant form' suggested by \z and Harrison more than two
decades earlier. (These tiny fluctuations in matter become galaxies
several billion years later, after
being amplified by gravitational instability.)
This milestone discovery was awarded the nobel prize
in physics in 2006.

An important property of the universe is that it is gravitationally
unstable. This means that small initial fluctuations in the density of matter
grow and become larger. A theoretical analysis of density fluctuations
had been carried out in the 1940's by the eminent Soviet physicist
Evgeny Lifshitz. Lifshitz had shown that in an expanding universe density
perturbations grow at the rather modest rate $\delta \propto t^{2/3}$.
This is very much slower than the exponentially rapid `Jeans instability',
$\delta \propto \exp{(\sqrt{4\pi G\rho} t)}$ which occurs in a static
universe. The reason for the difference
 is that cosmic expansion moves particles away from
one another while gravity pushes them together. Since the two influences
oppose each other, gravitational instability becomes weaker
if the universe expands. Although Lifshitz's treatment was rigorous
it had a fundamental limitation. In order to solve the complicated equations
of general relativity Lifshitz had to assume that perturbations were linear,
\ie $\delta \ll 1$, where $\delta = (\rho - {\bar\rho})/{\bar\rho}$ is the 
perturbation in the cosmic density relative to its mean value ${\bar\rho}$.
Although this approach was exceedingly useful in the context of the early
universe which was quasi-homogeneous, %so that $\delta \ll 1$,
it broke down at more recent times since density perturbations at
the present epoch are exceedingly large, the density contrast
associated with a galaxy being close to a million ($\delta \sim 10^6$).
%Though useful this approach was incomplete, since
%the density contrast in galaxies is close to
%Thus one could not use this approach to study galaxies which 
%had a density contrast of close to a million ($\delta \sim 10^6$).
\z set about rectifying this situation by proposing a remarkably simple and elegant
approximation which could be used to follow a perturbation
from its initially linear form into the fully
nonlinear regime when $\delta \gg 1$.
%His work led to the notion 
%arranging matter into cellular or honeycomb structure nowadays
%referred to as the {\em Cosmic Web}.
%In doing so 
Along the way \z upset a widely prevailing world view according to which
%Up until 1970 
%it was widely believed that the 
the assembly of the first large astrophysical objects in the universe was
spherical in nature. According to this point of view, the spherical globular
clusters that orbit our galaxy were the first objects
to condense out of an expanding quasi-homogeneous gas of neutral hydrogen.
\z toppled this deeply ingrained notion by 
demonstrating 
%(using an elegant formulation now known as the {\em Zeldovich approximation}),
 that gravitational instability was much more likely to proceed in
an anisotropic manner resulting in the formation of two dimensional sheet-like
objects which he called `Pancakes' %$^{10}$\footnotetext[10]{%The pancake
%after a delicious Russian dish
%is a translation of the Russian word `blin' which 
%is a popular and 
%tasty Russian dish. One could equally substitute pizza,
%chapati or dosa, depending upon ones favourite cuisine !
$^{10}$\footnotetext[10]{Its rather surprising that, while
 the Zeldovich approximation plays a key role in 
our present understanding of the {\em Cosmic Web}, its reception by the academic community 
was initially lukewarm. 
Indeed, in the decade following its publication,
the `Zeldovich approximation' paper was cited
less than 30 times and that too, mostly by scientists from within the Soviet Union \cite{shan_sun}.}.

The \z approximation proposes that the final coordinate of a 
particle is related to its initial coordinate by the transformation \cite{zel_approx}
\beq
{\bf r} = {\bf q} + {\bf v({\bf q})} \delta_{\rm \ell} 
\label{za}
\eeq
where $\delta_{\rm \ell}$ is the density contrast predicted by {\em linear
theory} and ${\bf v({\bf q})}$ is the initial velocity field of a perturbation.$^4$
(${\bf v({\bf q})}=0$ if particles remain at rest with respect to the
background cosmic expansion, in which case ${\bf r} = {\bf q}$.) If the particle flow is irrotational then,
under some assumptions, one can relate the velocity field to the linearised
gravitational potential 
\beq
{\bf v({\bf q})} = \nabla\phi~.
\label{za1}
\eeq
 In turn
$\phi$ is related to the primordial density perturbation $\delta_{\rm \ell}$
and the mean density of matter in the universe, ${\bar\rho}$,
through the Poisson equation$^{11}$\footnotetext[11]{The {\em comoving coordinate} ${\bf r}$ is a convenient quantity
because, in the absence of perturbations, its value does not change as the
universe expands. It
is related to the 
{\em physical coordinate} ${\bf x}$ by means of the transformation
${\bf r} = {\bf x}/a(t)$. The \z approximation in physical coordinates
 has the form 
${\bf x} = a(t)\lbrack{\bf q} + {\bf v({\bf q})} \delta_{\rm \ell} \rbrack$.
${\bf r}$ and ${\bf q}$ respectively represent
 the Eulerian and Lagrangian coordinate of a particle in
fluid dynamics.}

\beq
\nabla^2\phi = 4\pi Ga^2{\bar\rho}\times\delta_\ell~.
\label{za3}
\eeq
We noted earlier that $\delta_\ell \propto t^{2/3}$. Introducing a new
time coordinate $T = \delta_\ell \propto t^{2/3}$ allows us to rewrite (\ref{za}) as
\beq
{\bf r} = {\bf q} + {\bf v({\bf q})}T~.% ~~{\bf v({\bf q})} = \nabla\phi~.
\label{za2}
\eeq
We have thus shown that the \z approximation is equivalent to the simple inertial motion of
particles !
An essential feature of inertial motion from random initial conditions is
the intersection of particle trajectories leading to the formation of singularities
in the density field. A similar effect is seen in the propogation
of light as it passes through a medium such as a plate of glass or water.
After passing through the plate 
neighboring light trajectories intersect to form caustics where the intensity of light
is exceedingly bright, see figure \ref{fig:caustics}.

\begin{figure*}
\centerline{ \psfig{figure=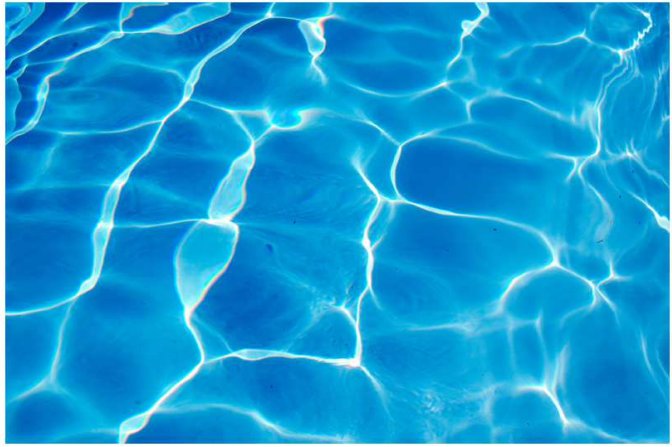,width=0.55\textwidth,angle=0} }
\bigskip
\caption{
Caustics of light in a pool of water.}
\label{fig:caustics}
\end{figure*}

Indeed, %it is quite remarkable that 
the \za bears a very simple
analogy to the propogation of light rays in geometrical optics !
Consider a light ray which enters a (two dimensional)
 glass plate at the point
${\bf q} = \lbrace q_1,q_2\rbrace$. If the thickness of the plate is 
$h = h(x,y)$ then, after passing through the plate, the light ray will be
deflected by an angle ${\vec\theta}({\bf q})$ which determines the direction of the ray
after it passes through the plate. The two dimensional coordinates of
the light ray after it has emerged from the plate will depend 
upon ${\vec\theta}$ as well as the location of the screen $z$, so that (see figure 1)
\beq
{\bf R}(z,{\bf q}) = {\bf q} + {\bf \theta}z~,
\label{optics}
\eeq
where 
\beq
\theta_i = -(n-1)\frac{\partial h({\bf q})}{\partial q_i}
\label{optics1}
\eeq
and $n$ is the refractive index of the plate. 

\begin{figure*}
\centerline{ \psfig{figure=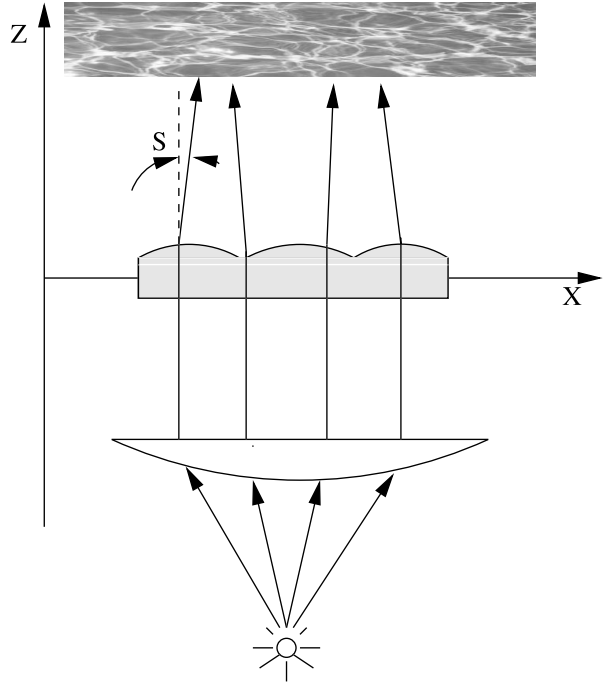,width=0.55\textwidth,angle=-90} }
\bigskip
\caption{
Caustics of light on a screen;
adapted from \cite{shan_zel}.}
\label{fig:caustics1}
\end{figure*}

A screen kept some distance away from the plate will see an
inhomogeneous distribution of light, see figure \ref{fig:caustics1}. Varying the location
of the screen one will see caustics, regions where light
trajectories have interesected causing the brightness of light to suddenly
shoot up. If we denote the brightness by light by $\rho$ then, it is easy to show that
\beq
\rho(z,{\bf q}) = \frac{\rho_0}{\lbrack 1- z\alpha({\bf q})\rbrack\lbrack
1-z\beta{(\bf q})\rbrack}~,
\label{optics2}
\eeq
where $\rho_0$ is the initial intensity of light and
 $\alpha({\bf q})$ and $\beta{(\bf q})$ are the principle curvatures of
the surface of the plate $h=h(\bf q)$. In other words $\alpha({\bf q})$ and $\beta{(\bf q})$ are
eigenvalues of the
tensor $\partial^2h/\partial q_i\partial q_k$. 
Clearly the optics relation
(\ref{optics}) is identical to the \za given by (\ref{za2}).
The similarity between optics and gravitational instability is striking ! 
The role of the plate thickness $h$ is played
by the gravitational potential $\phi$, and the location of the screen $z$
is analogous to the cosmic time coordinate $T$. 
In optics (gravitational instability) the intensity of light
(density of matter) becomes extremely large at caustics where nearby
light (matter) trajectories intersect. From (\ref{optics2}) we find
 $\rho \to \infty$
when $z \simeq \alpha^{-1}$ (assuming $\alpha > \beta$). A similar result
holds in the \za when we replace $z$ by $T$. 
In this case the law of matter conservation 
%$dM = \rho_0dV_L = \rho({\bf x},t) dV_E$ 
results in the following expression for
the density
\beq
 \rho({\bf x},t) = \frac{\rho_0}{\lbrack 1-T\alpha({\bf q})\rbrack
\lbrack 1-T\beta({\bf q})\rbrack \lbrack 1-T\gamma({\bf q})\rbrack}~, ~~
T \equiv \delta_\ell \propto t^{2/3}~,
\label{za4}
\eeq
where $\alpha, \beta, \gamma$ are the eigenvalues of the three dimensional 
deformation tensor $\partial^2\phi/\partial q_i\partial q_k$ and
$\phi$ is the inhomogeneous primordial gravitational potential responsible
for moving particles in (\ref{za}) \& (\ref{za2}).
Thus the \za predicts that matter moving under the influence of a perturbing
gravitational potential $\phi({\bf q})$ will get focussed into caustics,
%where its density will formally reach infinity (in practice pressure effects
%will prevent this from happening). The locations of these caustics is given
at locations specified by the eigenvalues of 
 $\partial^2\phi/\partial q_i\partial q_k$.
Whether a given volume element contracts or expands depends upon the
specific values of the eigenvalues $\alpha({\bf q}), \beta({\bf q}), \gamma({\bf q})$,
in a given region of space and especially
on their sign. If, at the point ${\bf q}$ 
one of the eigenvalues, say $\alpha({\bf q})$, is positive, 
then, since $T \propto t^{2/3}$ is a monotonically increasing function of time,
 a time will come when $1-\alpha T = 0$. From (\ref{za4}) we find that the denominator in
this expression will vanish as the density of matter
 aquires very large (formally infinite) values. This signals the formation of 
 a caustic at ${\bf q}$ .
%If $\phi$ is distributed in the manner of a Gaussian random field then
%such a situation will occur in 
Such a region will be the birthplace of a {\em pancake}. Intersections of 
fully grown pancakes
will form filaments and intersections of filaments will result in clumps.
Matter will have a multistream flow within pancakes, moving along
them towards filaments, and then moving along filaments to converge into clumps.
%Although the \za breaks down soon after the formation of the first caustics
%the appearance of the cosmic web in the form of filaments intersecting at the
%location of clusters appears to bear out this method of understanding the
%formation of large scale structure in the universe.
%This is precisely what i
%The locations of clumps might correspond to clusters of galaxies.

%Furthermore, since the eigenvalues $\alpha({\bf q}), \beta({\bf q}), \gamma({\bf q})$
%can be negative as well as positive, 
The formation of caustics will be accompanied by
the formation of immense underdense regions, or voids, %where the density contrast
%will rapidly decrease as matter 
from which matter gets drained into 
pancakes and filaments.
Thus the \za predicts that, as the universe expands, the matter in it
becomes concentrated 
along planar and filamentary `superclusters', and that neighboring superclusters 
are separated by `voids' -- vast empty regions
virtually deviod of the presence of matter, see figure \ref{fig:web}. It is remarkable that precisely
such a  {\em supercluster-void}
network of galaxies, now commonly referred to as the {\em Cosmic Web}, has been
discovered by large galaxy surveys almost 30 years after it was first predicted
to exist by \z \cite{gott03}. 
The supercluster which currently holds the record for being the largest contiguous
object in the universe is close to a billion light years across and has been dubbed 
the {\em Sloan Great Wall} since it was discovered by the Sloan digital sky survey,
see figure \ref{fig:greatwalls}. 

\begin{figure*}
\centerline{ \psfig{figure=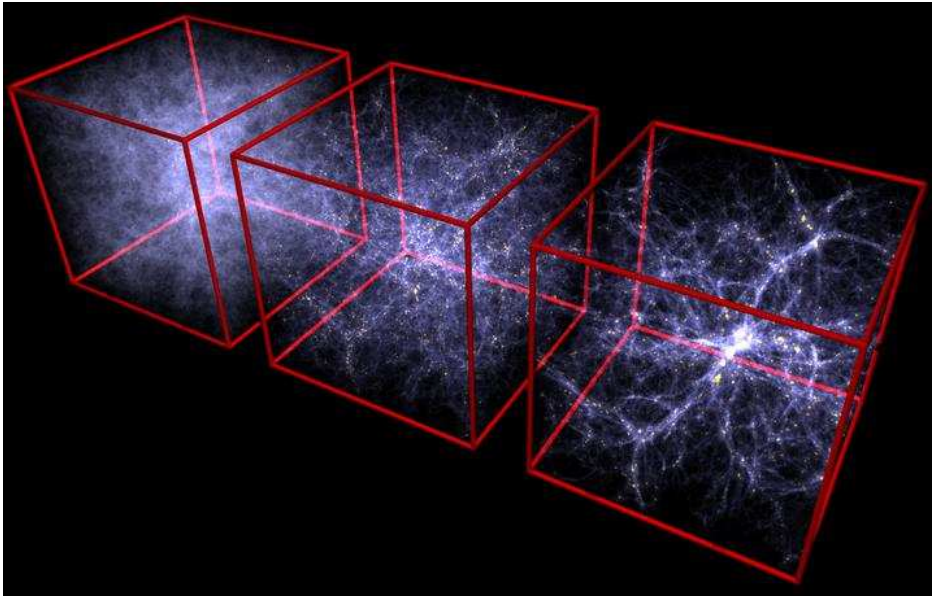,width=0.75\textwidth,angle=0} }
\bigskip
\caption{
The development of the {\em Cosmic Web} is shown 
      in a hydrodynamic N$-$body simulation from $z=6$
      (leftmost cube) to $z=0$ (rightmost cube) via $z=2$ (middle
      cube). A near featureless density field evolves to produce
      filamentary and sheet-like superclusters which percolate through
      the box-volume. These are separated by large voids. 
      [Figure courtesy: Volker Springel; see {\it
        http://www.mpa-garching.mpg.de/galform/data\_vis/index.shtml}]}
\label{fig:web}
\end{figure*}

\begin{figure*}
\centerline{ \psfig{figure=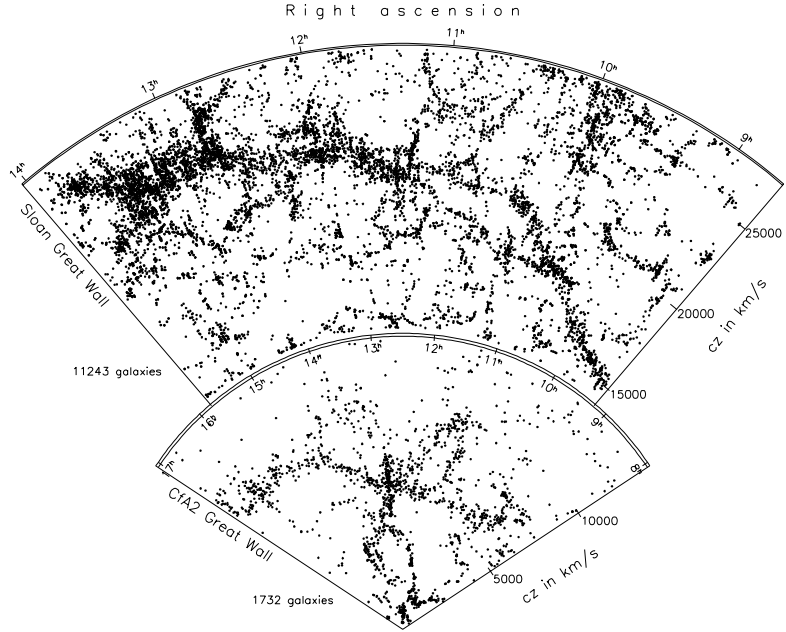,width=0.75\textwidth,angle=0} }
\bigskip
\caption{The {\em Sloan Great Wall} situated at $z\simeq0.08$ is shown
      in the upper panel of the figure. For comparison, a relatively
      nearer structure known as the CfA great wall is also shown. The linear
      extent of the Sloan Great Wall is about 500 Megaparsec, which is
      roughly twice as large as the CfA great wall. Figure courtesy of
      \cite{gott03}.}
    \label{fig:greatwalls}
\end{figure*}

%It is not clear whether this represents the largest conglamoration of matter 
%in the universe 
%or whether even larger superclusters and voids will be discovered by deeper survey's.
%Cosmology is full of surprises and one can anticipate many exciting developments
%once the cosmic web is mapped with upto a billion galaxies as promised by future survey's.
%such as SKA (the Square Kilometer Array).
%in great detail in the coming decades with survey's which promise to map the positions
%of upto a billion galaxies !
%characterizes the distribution of matter in the universe on the very largest scales

In his later years, \z, together with students Doroshkevich
and Shandarin and the eminent Soviet mathematician Vladimir Arnold,
developed a rigorous mathematical understanding of gravitational clustering
in the universe involving sophisticated mathemetical methods
which included catastrophy theory and percolation analysis.
Work in this direction continues to this day \cite{sahni_coles}.

\item The {\em Cosmic Web} is spectacular.
The `atoms' of this {\em web} are galaxies, which gravitationally
bind together
to form clusters of galaxies (a `rich' cluster can contain several
thousand galaxies)
and the much larger superclusters \cite{cosmic_web}. %(A supercluster can contain upto
%a hundred thousand galaxies and span several hundred million light years.)
In 1972, \z and Sunyaev published a seminal paper in which 
they showed that photons from the cosmic microwave background would scatter
off the hot plasma trapped in the deep potential wells of clusters. This would alter the brightness
of the CMB when viewed in
the direction of a cluster. Since its prediction almost four decades ago, the
{\em Sunyaev-Zeldovich effect} has been observed in many galaxy clusters and promises
to pave the way for a deeper understanding of our universe on the very largest
scales, perhaps even throwing light on
 the elusive nature of dark energy \cite{biman}.

\end{itemize}

\section{Personal Reminiscences}

\centerline{``Glaza ne vidyat a ruki delayut '' -- Ya.B. Zeldovich}
\centerline{\em Transl. ``Although the eyes don't see the hands continue creating~$^{12}$.\footnotetext[12]{
This is very similar to Jung's:
``Often the hands will solve a mystery that the intellect has struggled with in vain''.}''}

\medskip

My first meeting with \z took place in 1978 when, as a student
of physics at Moscow State University, I was looking for a professor who
would guide me for my (pre-MSc) course work. \z agreed and, to test my skills,
gave me a project in general relativity, a subject which I had just started to learn.
I soon realized that \z was an excellent teacher and could explain in very simple
language exceedingly complex physical ideas. This was a great boon to students
since, despite his extremely busy scientific schedule, \z always found time
to teach courses at
Moscow University. So it was that I was initiated into
the intricate and beautiful field of cosmology via a twenty two lecture course taught by
\z -- each lecture being of roughly 2 hour duration with a ten minute break in between. Particularly impressionable for us youngsters was the fact that though 
teaching a  mathematically rigorous course, \z always took great pains to explain
even the most difficult ideas using 
simple physical arguments and intution.  (\z did this even at regular seminars. 
If one of the talks excited him then, when it ended, \z
would bound up to the podium and provide an impromptu summary.
His great breadth of interests allowed \z to summarize talks on virtually
any subject and I have seen him apply this skill to topics ranging from
stellar physics to particle physics and quantum gravity. 
I should add that these 2 minute summaries were invaluable to us students,
 since they were frequently more
lucid and transparent than the original seminar !)

\z continuously modified and expanded
his course material taking care to ensure that significantly new developments
in the field were covered. (On attending the
very same cosmology course 4 years later, I was pleasantly surprised at finding
that almost a quarter of its content was new. \z assiduously incorporated several
 recent
developments including grand unification, topological
defects and inflation, into his `self-revised' syllabus.)
Participants at his lectures consisted not only of students, but also
senior researchers and professors, many of whom stayed back after class
to discuss new science ideas with \z. 
Another remarkable quality of \z was his willingness to learn from others
and to acknowledge, often in public, the mistakes which he had made, and what
could be learned from them. Thus he admitted in class how he had misunderstood the
data regarding the cosmic microwave background in the early 1960's and,
to his great regret, had initially advocated the cold big bang model 
instead of the hot one (the latter had been predicted by Gamow
and subsequently shown to be correct).
%I realised how rare a trait this was when, years later, I was to meet
%other eminent scientists who doggedly defended their own favourite viewpoint
%even in the face of mounting evidence to the contrary.

\z 
strove to explain complicated ideas simply through numerous entertaining
articles and text books \cite{zel0,zel00}
including his excellent monograph `Higher Mathematics for Beginners'
\cite{zel_maths} which presented
serious mathematical ideas in a form which is accessible to a high school student.
He once wrote ``the so-called `strict' proofs and definitions are far more
complicated than the intuitive approach to derivatives and integrals. 
As a result, the mathematical ideas necessary for an understanding of physics
reach school-pupils too late. It is like serving the salt and pepper not for lunch,
but later -- for afternoon tea''.
In this \z was following an old physics tradition epitomized by Landau in Russia and
Einstein and Feynman in the west, 
of bringing the excitement of
science closer to students and `ordinary' people.

Zeldovich's manner of conducting exams was also quite unusual !
When, towards the end of his cosmology course the time came for exams, I realised 
that there was no formal time table for a written test, as was usually the case
$^{13}$\footnotetext[13]{Although Zledovich's class had been full,
only one other student apart from me decided to take the 
cosmology exam. The reason, as I came to know later, was Zeldovich's
rather high standards which ensured that most people failed the first time around.}.
%When I persisted with my desire to take an exam
Instead \z asked me to meet him outside his office a few days later.
When I did, \z scribbled two problems and asked me to solve them, after which
he walked away.
Although I had gone through his course material diligently 
I could not, even after trying hard, figure out how either
of these problems could be solved, and so feeling quite dejected I walked back
to my hostel a few miles away. This was summer 1979. By winter, with the heavy snows
of Moscow having set in, I 
managed to crack
one of the problems and, feeling rather elated, went back to \z.
\z subjected me to a strenuous viva-voce after which he declared that I had passed
with an A and, even more significantly, that my solution to his problem
was new and original and could be published as a paper $^{14}$\footnotetext[14]{\z thought
highly of the prospects of Indian science and encouraged me to publish my
results in an Indian journal. Thus my first paper was published in 
Pramana in 1980, considerably before I had completed my MSc.}. 

Although most people found \z to be very inspiring, some of his colleagues 
found his intellectual brilliance rather intimidating.
Indeed \z did not `suffer fools lightly' and I have seen him demolish
in less than a few minutes
many a senior scientist propounding a silly idea.
(At the same time \z was much more tolerant of students.
I occasionally blundered in his presence only to see him smile
and explain, with much patience and goodwill, just where I had gone wrong.)
\z set very high standards both for himself and the group of students
and collaborator's who made up his school or {\em gharana}. In so doing he
taught and nurtured a whole generation of talented Russian scientists
many of whom developed, in the course of his tutelage,
 diametrically complementary styles ranging from the
highly intuitive to the rigorously mathematical.
$^{15}$\footnotetext[15]{Sakharov mentions that Zeldovich's
`` effect on his pupils was remarkable; he often discovered in them a
capacity for scientific creativity which without him would not have been realized
or could have been realized only in part and with great difficulty.''}
 In this he was following
the tradition of the other great Russian {\em ustaad}, Landau, who like \z,
fostered and left behind a great scientific legacy in the form of a {\em gharana
of physics} which was almost unique in scientific method and style.

I experienced \z empathy on many occasions. 
Some years after commencing work on my PhD I was shocked to hear %I had 
%another experience which again highlighted his essential humanity and goodwill.
that \z's wife had just died of a serious illness. 
Together with some friends I went to pay my respects and offer condolences.
Even before I could utter a word, \z turned to me and offered his own
profuse condolences on the assasination of Indira Gandhi who had lost her
life to extremists the very same day that \z's wife lost hers to illness. 
In his eyes I could see how deeply
he felt, and I was profoundly moved that he could place the historical
anguish of a neighboring
 nation on the same footing as his own very deep  and personal loss.

Although very fond of travel, 
\z faced with numerous travel restrictions.
due to his early involvement with
the Soviet defence program.
In 1982, perhaps because of my political innocence and youthful enthusiasm,
I was very keen that \z visit India, and before embarking home for holidays
asked him what I could do to ensure his visit. Looking at me wistfully
\z remarked, ``they (the soviet authoroties)
will reply to your governments invitation saying
\z is ill and unable to travel, but look at me (\z flexed his muscles)
I am perfectly fit and can travel tomorrow !''.
Indeed \z's visit to India never did materialise and the following year
I bore witness to the very bizarre policies of the Soviet government
with regard to foreign travel by its eminent scientists.
\z and Starobinsky had both been
 invited to travel to a famous meeting on general relativity
(GR10) scheduled in Padova, Italy in 1983. Indeed, \z had been invited to organise a
special session on the Early Universe
$^{16}$\footnotetext[16]{The Early Universe had emerged as an important new field in the 1980's
whose promising directions included: Inflation, 
cosmological consequences of grand unified theories including 
baryogenesis and leptogenesis, 
particle physics candidates for dark matter,
quantum-cosmology and the wave-function of the universe, etc.}
%\z, being a world leader in this field, was a natural choice for session
%organiser.}
 while Starobinsky had consented to give a plenary
talk on Inflation, which was then a very hot topic. I too was planning to go 
since my paper (written jointly with Starobinsky and another of his pupils Lev Kofman)
had been accepted for presentation. I was very excited by the prospect
of travelling to Italy from Moscow (3 days journey by train passing through several
countries) and the day before
leaving I met Starobinsky who took me to \z with the remark
``Yakov Borisovich here is Varun, the only member of our delegation who is
sure of travelling to Italy'' ! I was not a little surprised to learn
that, with only four days remaining for the meeting, neither \z nor Starobinsky had yet
received official sanction for their visit.

I was broken hearted when, on reaching Padova 4 days later, I learned that
neither of my guru's had been allowed to travel, and that the
organisers were in a quandary as to how to salvage the early universe session.
%(Eventually Jim Hartle, an eminent american relativist, chaired it.)
What was even more disheartening was the fact that while neither \z nor Starobinsky
had been allowed to make the journey to Italy
(and showcase Soviet science), the official Soviet delegation was nevertheless replete
with scientists, of rather mediocre quality but with deep party connections,
who despite their ideological leanings were not ashamed to thoroughly enjoy
 `capitalist
hospitality'.

\z was finally allowed to travel outside of the Soviet
block in 1982 when, at the age of 68, he delivered an invited lecture 
``Remarks on the Structure of the Universe'' to the International Astronomical
Union in Patras, Greece. When asked by Jerry Ostriker, an eminent american 
astrophysicist, when he was last out of the Soviet Union, \z unhesitatingly
answered ``sixty eight years ago'', \ie in a previous life.

Viewed in retrospect, and with the demise of the Soviet Union, it appears
quite incomprehensible as to why many of its leading scientists could not
travel abroad and enjoy an envigorating science discussion with their contemporaries
in the east and west. 
In \z's case this was all the more surprising since his work had gained
widespread international recognition and \z was a foreign member both of the
Royal Society of London and of the U.S. National Academy
of Sciences. Nevertheless, if the mountain cannot come to Muhammed then
Muhammed must go to the mountain,
and so the best minds from Europe
and the U.S. travelled to the Soviet Union to meet with \z 
and interact with the other
eminent scientists belonging to 
the `Moscow school'.
So it was that I got the opportunity of listening
 to inspiring lectures by Stephen Hawking and Subrahmanyan
Chandrasekhar at special seminars organised by \z at Moscow's
Sternberg astronomical institute. %and with his inspiring
%commentary afterwards !

It was refreshing to see that, despite his remarkable talents,
\z never pushed ideas simply because they were his own.
I was a witness to this when, on one occasion, after a seminar had
formally ended \z called us back into the auditorium saying there was
a `small idea' which he wanted to discuss. The idea turned out to be
that of cosmic strings which, having been created during phase transitions in
the early universe could, 
rather remarkably, seed galaxy
formation. This was in 1980 and \z had just published
a paper in the Monthly Notices of the Royal Astronomical Society describing
his scenario. However, this was also the period when the Inflationary scenario was
propounded (independently by Alan Guth in the US and Alexei
 Starobinsky 
in Moscow). \z was fascinated by inflation, and even though he was not its
discoverer, he championed this model enthusiastically \cite{zel1}.
When, during 1981--1982, it became clear that the rapidly expanding inflationary
universe could quantum-mechanically generate the perturbations required to seed
galaxies, \z saw the logical beauty of this approach and abandoned, in the process,
his own string-based galaxy formation scenario. 
Perhaps because of this topological defects
never did gain popularity in the Soviet Union, while interest in them
 thrived for more than
two decades in the west (partly due to the influence of the talented Ukrainian
emigr\'e Alex Vilenkin).
%His reluctance to work on strings

Although \z worked undoubtedly hard, he was hardly a nerd and thoroughly
enjoyed both entertainment and sport (he had boxed in his youth and loved
to swim and ski). He was also very well versed in literature and frequently quoted 
from writers and poets while lecturing on physics $^{17}$\footnotetext[17]{Perhaps \z had imbibed
his love of prose and poetry from his mother, who had been a translator of literature
from French into Russian.
In this I felt a certain affinity with \z since my father too was a
litt\'erateur and, during the 1960's, had translated several Russian classics
into Hindi.}. His sense of humour was also legendary and he used it abundantly
in class, which made his lectures far from boring !

I hope I have been able, within the short span of this article, to communicate
to readers of Resonance what a remarkable man Yakov Borisovich \z really was. His deep dedication
to science combined with great originality and a contagious enthusiasm made him one of
the most influential theoretical physicists of the last century.
I would like to end by quoting from Andrei Sakharov who was a long time
colleague and friend of \z \cite{sakharov}:
``Now, when Yakov Borisovich Zeldovich has departed from us, we, his friends and
colleagues in science, understand how much he himself did, and how much he gave
to those who had the chance to share his life and work''.

\section*{Acknowledgments}
I thank Biman Nath and Anvar Shukurov for stimulating conversations.

\end{document}